\documentclass{ws-p8-50x6-00}

\begin{document}

\title{ Fractional Fokker-Planck Equation in Time Variable 
   and Oscillation of Cumulant Moments  }
\author{N. Suzuki}

\address{Matsusho Gakuen Junior College, Matsumoto 390-1295, Japan
  \\E-mail: suzuki@matsu.ac.jp} 

\author{M. Biyajima}

\address{Department of Physics, Shinshu University, Matsumoto 390-8621
Japan}

\maketitle

\abstracts{ Fractional derivative in time variable is introduced into the Fokker-Planck equation of a population growth model.  It's solution, the KNO scaling function, is transformed into the generating function for the multiplicity distribution.  Formulas of the factorial moment and the $H_j$ moment are derived from the generating function, which reduces to that of the negative binomial distribution (NBD), if the fractional derivative is replaced to the ordinary one.  In our approach, oscillation of $H_j$ moment appears contrary to the case of the NBD.  Calculated $H_j$ moments are compared with those given from the data in $p\bar{p}$ collisions and in $e^+e^-$ collisions.}

\section{Introduction} 
   The negative binomial distribution (NBD) is often used for the analysis of observed multiplicity distributions in high energy hadron-hadron ($hh$), and $e^+e^-$ collisions. The $H_j$ moment derived from the generating function of the NBD does not show oscillatory behavior as the rank of cumulant moment increases.  On the othe hand, $H_j$ moments obtained from observed multiplicity distributions in $hh$ and $e^+e^-$ collisions show oscillatory behaviors~\cite{drem93}.
If the truncated NBD is used to calculate the $H_j$ moments in $hh$ collisions, calculated results can describe those obtained from the data well~\cite{naka96,ugoc95}. However, in $e^+e^-$ collisions, calculated $H_j$ moments by the use of the truncated NBD oscillate much weaker than those obtained from the data, and cannot explain the oscillatory behavior of the $H_j$ moments obtained from the data~\cite{suzu96}.

  The NBD is derived from a branching process, birth and death process with immigration. The probability distribution (the multiplicity distribution) and the branching equation are transformed respectively into the probability density  (the KNO scaling function) and the Fokker-Planck equation by the Poisson transform.   

The fractional calculus has a long history~\cite{oldh74}.  Recently, the fractional Fokker-Planck equation have been introduced to describe anomalous diffusion phenomena~\cite{wyss86,carp97,bark00}; the fractional space or time derivative replaces the ordinary space or time variable.  If fractional derivative in time variable is introduced to the Fokker-Planck equation, it means that a sort of time lag or a memory effect is introduced into the Fokker-Planck equation, or the branching equation.  We would like to investigate this type of fractional Fokker-Planck equation, the solution of which reduces to the gamma distribution when the fractional derivative is replaced to the ordinary one, and an effect of fractional derivative on the behavior of cumulant moments.

\section{The Fractional Fokker-Planck Equation}

In the following we investigate the fractional Fokker-Planck equation (FFPE), 
 \begin{eqnarray}
  \frac{\partial \psi (z,t)}{\partial t}
                &=& {}_0{\rm D}^{1-\alpha}_tL_{FP} \psi (z,t), 
                 \quad 0<\alpha<1, \nonumber \\
  L_{FP} &=& -\frac{\partial}{\partial z}
    \left[ a(z) - \frac{1}{2}\frac{\partial}{\partial z}b(z) \right],
    \quad a(z)= \beta - \gamma z,  \quad b(z)= \sigma^2 z,    
     \label{eq.frac1} 
 \end{eqnarray}
with the initial condition,
 \begin{eqnarray}
   \psi (z,t=0) = \delta(z-z_0) \label{eq.frac2}.
 \end{eqnarray}
In Eq.({\ref{eq.frac1}),  $\beta>0$, $\sigma>0$ and $\gamma$ is real.  Symbol ${}_0{\bf D}_t^\delta$ denotes the Riemann-Liouville fractional derivative~\cite{oldh74} defined by
 \begin{eqnarray}
   {}_0{\bf D}_t^\delta f(t) = \frac{1}{\Gamma(n-\delta)} \frac{d^n}{dt^n}
             \int^t_0 (t-\tau)^{n-\delta-1} f(\tau) d\tau, 
   \quad n-1 \le \delta <n,  \label{eq.frac3}
 \end{eqnarray}
where $n$ is a natural number.  If $\alpha=1$, ${}_0{\bf D}_t^0 f(t)=f(t)$ ( ${}_0{\bf D}_t^0 =1$ ), and Eq.(\ref{eq.frac1}) reduces to the ordinary Fokker-Planck equation.

According to the method proposed by Barkai and Silbey~\cite{bark00}, 
we assume that
 \begin{eqnarray}
       \psi(z,t) = \int_0^{\infty}  R_s(t) G_s(z) ds,   \label{eq.frac4}
 \end{eqnarray}
and that function $G_s(z)$ satisfies the following equations,
 \begin{eqnarray}
    L_{FP} G_s(z) = \frac{\partial}{\partial s} G_s(z), \quad 
    G_0(z)=\delta(z-z_0).  \label{eq.frac5}
 \end{eqnarray}
Then function $G_s(z)$ becomes the KNO scaling function for the Pe\v{r}ina-McGill formula,
 \begin{eqnarray}
    G_s(z)&=& \frac{1}{kp} 
          \exp[-\frac{z+z_0(1-p)}{kp}]
          \left( \frac{z}{z_0(1-p)} \right)^{(\lambda -1)/2}
      I_{\lambda-1} ( \frac{2 \sqrt{zz_0(1-p)} }{kp} ),
                  \nonumber \\
     k &=& \frac{\sigma^2}{2\gamma},  \quad
    \lambda = \frac{2 \beta}{\sigma^2}, \quad
     p = 1 - e^{-\gamma s}.                    \label{eq.frac6}
 \end{eqnarray}

Applying the Laplace transform to Eq.(\ref{eq.frac1}), we find
 \begin{eqnarray}
     \int_0^\infty \left[ u \tilde R_s(u)
        +  u^{1-\alpha}\frac{\partial \tilde R_s(u)}{\partial s} 
            \right] G_s(z) ds
       = \left[ 1 - u^{1-\alpha}\tilde R_0(u) \right]\delta(z-z_0), 
         \label{eq.frac8}
 \end{eqnarray}
where $\tilde R_s(u)$ is the Laplace transform of $R_s(t)$,
 \begin{eqnarray}
   \tilde R_s(u) = \int_0^{\infty} R_s(t) e^{-ut} dt. 
                 \label{eq.frac9}
 \end{eqnarray}
Furthermore, we assume that each side of Eq.(\ref{eq.frac8}) is equal to zero;
 \begin{eqnarray}
    u^{1-\alpha}\tilde R_0(u) = 1,  \quad 
    -u^{1-\alpha} \frac{\partial \tilde R_s(u)}{\partial u} = u\tilde R_s(u).
               \label{eq.frac10}
 \end{eqnarray}
The solution of Eq.(\ref{eq.frac10}) is given by
 $ \tilde R_s(u) = u^{\alpha-1} \exp[-su^\alpha].  $
Then $R_s(t)$ is written as
 \begin{eqnarray}
   R_s(t) = \frac{t^{-\alpha}}{2\pi i}
              \int_{c_0-i\infty}^{c_0+i\infty}\sigma^{\alpha-1}
             \exp[\sigma - \frac{s}{t^\alpha} \sigma^\alpha] d \sigma,
                \quad (c_0>0).
                \label{eq.frac11}
 \end{eqnarray}

In the following, we assume that $\gamma>0$. Equation (\ref{eq.frac6}) 
can be expanded as
 \begin{eqnarray}
    G_s(z)= \frac{1}{k} ( \frac{z}{k} )^{\lambda -1} 
           \exp[-\frac{z}{k}]
           \sum_{m=0}^{\infty} \frac{m!}{\Gamma(m+\lambda)}
           L_m^{(\lambda-1)}(\frac{z}{k})  L_m^{(\lambda-1)}(\frac{z_0}{k}) 
           e^{-m\gamma s},  \label{eq.frac12}
 \end{eqnarray}
where $L_m^{(\lambda-1)}(z)$ denotes the Laguerre polynomial.  
Then the solution for the FFPE (\ref{eq.frac1}) with $z_0=+0$ is given by
 \begin{eqnarray}
    \psi(z,t) = \frac{z^{\lambda -1}}{k^{\lambda}}  
           \exp[-\frac{z}{k}]
           \sum_{m=0}^{\infty} \frac{m!}{\Gamma(m+\lambda)}
           L_m^{(\lambda-1)}(\frac{z}{k}) E_{\alpha}(-m\gamma t^\alpha),
           \label{eq.frac13}
 \end{eqnarray}
where $E_\alpha(-t)$ for $t>0$ denotes the Mittag-Leffler function~\cite{carp97},
 \begin{eqnarray}
    E_\alpha(-t) = \frac{\sin(\alpha\pi)}{\alpha\pi} \int_0^\infty
           \exp[-(xt)^{1/\alpha}] \frac{1}{x^2+2x\cos(\alpha\pi)+1}dx.
                           \label{eq.frac14}
 \end{eqnarray}
It is written in the infinite series as
 \begin{eqnarray}
    E_\alpha(z) = \sum_{n=0}^{\infty} \frac{z^n}{\Gamma(\alpha n+1)}.
                          \label{eq.frac15}
 \end{eqnarray}
if $\alpha=1$, Eq.(\ref{eq.frac13}) reduces to the gamma distribution, the KNO scaling function of the NBD.

\section{Factorial Moment and Cumulant Moment}

The generating function for the multiplicity distribution $P(n,t)$ is defined as
 \begin{eqnarray}
    \Pi(u) = \sum_{n=0}^{\infty} P(n,t) u^n.  \label{eq.fact1}
 \end{eqnarray}
The multiplicity distribution $P(n,t)$ is connected to it's KNO scaling function $\psi (z,t)$ by the Poisson transform,
 \begin{eqnarray}
   P(n,t) = \frac{\langle n_0 \rangle ^n}{n!}
             \int^\infty_0 z^n \exp[-\langle n_0 \rangle z] \psi(z,t) dz
      \label{eq.fact2}.
 \end{eqnarray}
Then the generating function (GF) is written as
 \begin{eqnarray}
    \Pi( 1-{u}/{\langle n_0 \rangle} )
        = \int_0^\infty \psi(z,t) e^{-uz}dz.
                         \label{eq.fact3}
 \end{eqnarray}
The GF corresponding to Eq.(\ref{eq.frac13}) is given by
 \begin{eqnarray}
    \Pi(u) = \sum_{m=0}^{\infty} 
       \frac{\Gamma(m+\lambda)}{m!\Gamma(\lambda)}
       \frac{ \left( -k\langle n_0 \rangle (u-1) \right)^m }
         {\left(1 -k\langle n_0 \rangle (u-1) \right)^{m+\lambda}}
            E_{\alpha}(-m\gamma t^\alpha).
                       \label{eq.fact4}
 \end{eqnarray}
The factorial moment are defined as
 \begin{eqnarray}
    f_j = \frac{\partial^j \Pi(u)}{\partial u^j}\Big|_{u=1}
        = \langle n(n-1)\cdots (n-j+1)\rangle .  
                     \label{eq.fact5}
 \end{eqnarray}
The normalized factorial moment is given from Eqs.(\ref{eq.fact4}) and (\ref{eq.fact5}) as
 \begin{eqnarray}
    F_j =\frac{f_j}{\langle n \rangle^j}
        = \frac{\Gamma(\lambda+j)}{\Gamma(\lambda)\lambda^j}
    \frac{\sum_{m=0}^{j}(-1)^m{}_jC_m 
          E_{\alpha}(-m\gamma t^\alpha)}
       {[1-E_\alpha(-\gamma t^\alpha)]^j}.   \label{eq.fact6}
 \end{eqnarray}
The $k$th rank cumulant moment is defined by the following equation,
 \begin{eqnarray}
    \kappa_j = {\partial^j \ln\Pi(u)}/{\partial u^j}|_{u=1}.
                  \label{eq.fact7}
 \end{eqnarray}
Then we obtain the recurrence equation for the $H_j$ moment,
 \begin{eqnarray}
      H_1 = 1,  \quad
      H_j = 1 - \sum_{m=1}^{j-1}{}_{j-1}C_{m-1}\frac{F_{j-m}F_m}{F_j}H_m, 
                                        \label{eq.fact8}
 \end{eqnarray}
where    $H_j = {\kappa_j}/{f_j}$.  

If $\alpha=1$, Eq.(\ref{eq.fact3}) reduces to the GF for the NBD,
 with mean multiplicity $\langle n \rangle=k\lambda\langle n_0 \rangle(1-e^{-\gamma t})$,
\begin{eqnarray}
       \Pi_{\rm NB}(u) = [ 
         1- ({\langle n \rangle}/{\lambda})(u-1)
                    ]^{-\lambda}.        \label{eq.fact9}
\end{eqnarray}
The normalized factorial moment and the $H_j$ moment for the NBD are given from Eq.(\ref{eq.fact9}) respectively as,
\begin{eqnarray}
       F_{\rm NB},_j = \frac{\Gamma(\lambda+j)}
                      {\Gamma(\lambda)\lambda^j}, \quad
       H_{\rm NB},_j = \frac{\Gamma(\lambda+1)(j-1)!}{\Gamma(\lambda+j)}.
      \label{eq.fact10}
\end{eqnarray}
As can be seen from Eqs.(\ref{eq.fact6}) and (\ref{eq.fact10}), difference between the normalized factorial moment derived from the FFPE ($0<\alpha<1$)  and that of the NBD ($\alpha=1$) is given by Mittag-Leffler functions.
\section{Calculated Results and Discussions}

In order to calculate $H_j$ moments, we use observed values of 
$\langle n \rangle$ and $C^2 (=\langle n^2 \rangle/\langle n \rangle^2)$ for the charged particles. Then, if $\alpha$ and $\gamma t^\alpha$ are  given, $\lambda$ in Eq.(\ref{eq.fact6}) is determined by the following equation, 
 \begin{eqnarray*}
    \frac{1}{\lambda}
        = \left(C^2 - \frac{1}{\langle n \rangle} \right)
          \frac{[1-E_\alpha(-\gamma t^\alpha)]^2}
               {1-2E_\alpha(-\gamma t^\alpha) + E_{\alpha}(-2\gamma t^\alpha)}
          -1.
                         \label{eq.calc1}
 \end{eqnarray*}

In the following, calculated results with $\alpha=0.5$ are shown. 
 In Fig.1a, our results with $\gamma t^\alpha=1.0$ ($\lambda=$66.82) and 2.0 (10.24) are compared with the $H_j$ moment obtained from the data in $p\bar{p}$ collisions at $\sqrt{s}=546$GeV~\cite{ua587}.  We can see that calculated values with $\alpha=0.5$ and $\gamma t^\alpha=1.0$ oscillate as strong as those from the data.  
If parameter $\gamma t^\alpha$ is adjusted with a step of 0.01 so that the first relative minimum of the calculated $H_j$ moment should be located at $j\ge 5$, 
the calculated first relative minimum value is $H_7=-3.323\times 10^{-5}$, and the absolute value of it is much smaller than that obtained from the data.  

 In Fig.1b, the calculated $H_j$ moment is compared with that obtained from the data in $e^+e^-$ collisions at $\sqrt{s}=91$ GeV~\cite{opal92}.  Parameter $\gamma t^\alpha$ is adjusted so that the first relative minimum should be located at $j\ge5$. As can be seen from the figure, the strength of the oscillation of calculated $H_j$ moment is comparable with the data.
  Even if the truncated NBD is used to calculate the $H_j$ moments, the oscillation of calculated moments is much weaker than that of the data~\cite{suzu96}.  Therefore, we can improve the results much better by introducing fractional derivative into time variable of the Fokker-Planck equation.

Calculated results from the GF with $\alpha=$0.5 show that the oscillation of $H_j$ moments are comparable with the data in amplitude both in $p\bar{p}$ and $e^+e^-$ collisions.  However, the first relative minimum points, or periods of oscillations are not necessarily the same as the data. 

The fractional moments have been also investigated by Hegyi~\cite{hegy96}. However, the formula derived from Eq.(\ref{eq.frac1}) is different from those in ref.~\cite{hegy96}.  

\begin{figure}[t]
 \begin{center}
 \begin{minipage}{0.47\linewidth}
  \begin{center}
   \epsfxsize=10pc
   \epsfbox{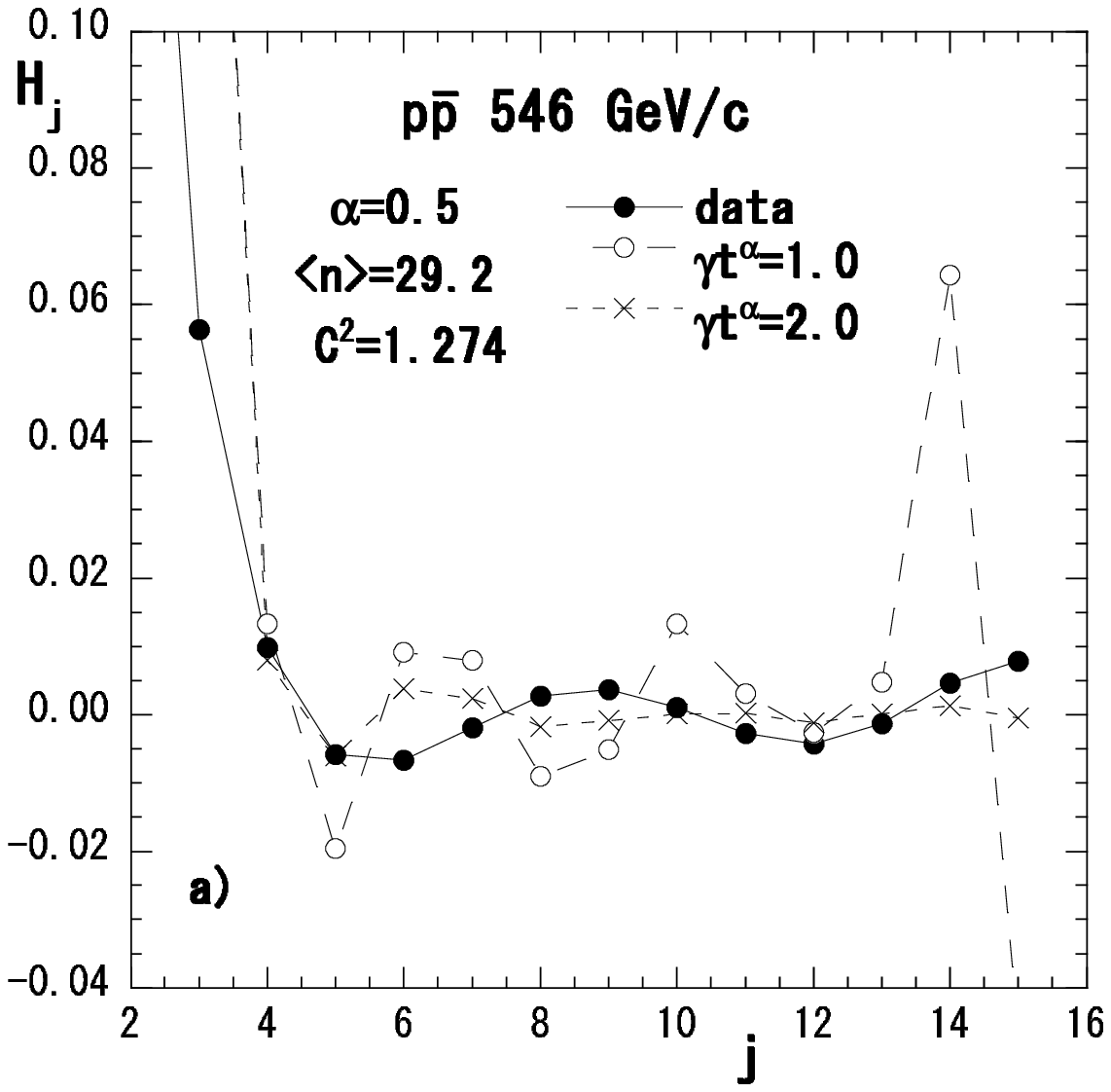}
    \end{center}
 \end{minipage}
 \begin{minipage}{0.47\linewidth}
  \begin{center}
   \epsfxsize=10pc
   \epsfbox{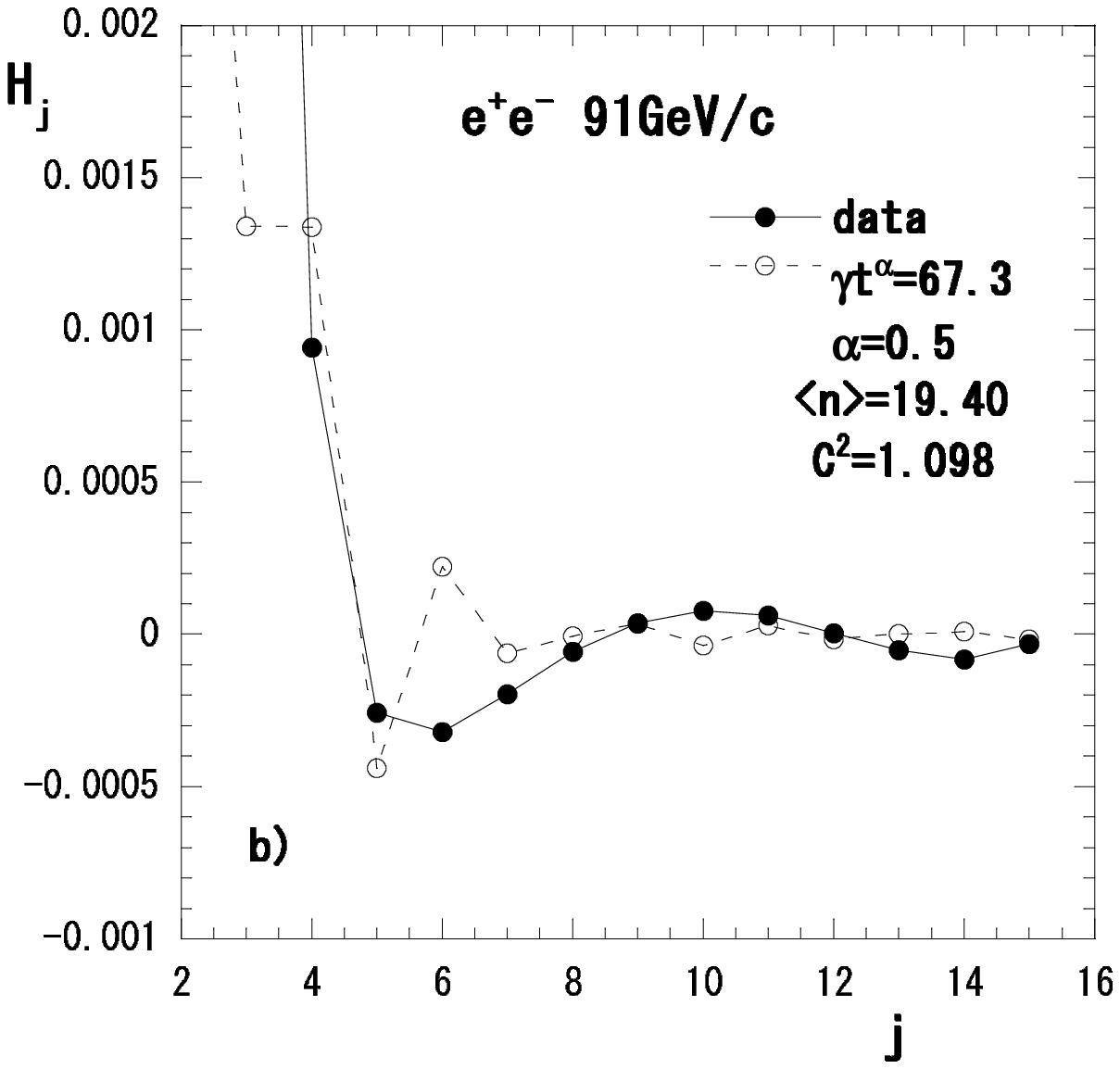}
   \end{center}
 \end{minipage}
\caption{ Calculated $H_j$ moments as a function of rank $j$ are 
compared with $H_j$ moments obtained from the data, 
a) in $p\bar{p}$ collisions, and b) in $e^+e^-$ collisions.}
\end{center}
\end{figure}


\begin{thebibliography}{99}
%
%
 \bibitem{drem93} I. M. Dremin and V. A. Nechitailo, JETP
                  Lett. {\bf 58}, 881 (1993); \\
                  I. M. Dremin and R. Hwa, Phys. Rev. {\bf
                  D49}, 5805 (1994);
                  I. M. Dremin, et al. Phys. Lett. {\bf B336}, 119 
                  (1994).
%
 \bibitem{naka96} N. Nakajima, M. Biyajima and N. Suzuki,
                   Phys. Rev. {\bf D54}, 4333 (1996).
%
 \bibitem{ugoc95} R. Ugoccioni, A. Giovannini and S. Lupia,
                  Phys. Lett. {\bf B342}, 387 (1995).
%
 \bibitem{suzu96} N. Suzuki, M. Biyajima and N. Nakajima,
                  Phys. Rev. {\bf D53}, 3582 (1996) and 
                  {\bf D54}, 3653 (1996).
%
 \bibitem{oldh74} K. B. Oldham and J. Spanier, The Fractional Calculus, Academic                  Press, New York, 1974; 
                 I. Podlubny, Fractional Differential Equations, Academic Press,                   San Diego, 1999.
%
 \bibitem{wyss86} W. Wyss, J. Math. Phys. {\bf 27}, 2782 (1986); 
                  W. R. Schneider and W. Wyss, J. Math. Phys. {\bf 30}, 134 (1989).
%
 \bibitem{carp97} A. Carpinteri and F. Mainardi, Fractional Calculus in Continuum Mechanics,                        Springer Verlag, Vienna, 1997.
%
 \bibitem{bark00} E. Barkai and R. J. Silbey, cond-mat/0002020.
%
%
 \bibitem{ua587} UA5 Collaboration, G. J. Alner et al., 
         Phys. Rep. {\bf 154}, 247 (1987).
%
 \bibitem{opal92} OPAL Collaboration, P. D. Acton et al., Z. Phys. {\bf C53}, 
      539 (1992).
%
\bibitem{hegy96} S.Hegyi, Phys. Lett. {\bf B387}, 642 (1996), 
                   and {\bf B417}, 186 (1998). 
\end{thebibliography}
\end{document}